\documentclass[twocolumn,english,aps,prl,showpacs]{revtex4}
\usepackage{graphicx}

\makeatletter

\usepackage{babel}
\makeatother
\begin{document}

\title{Finite Layer Thickness Stabilizes the Pfaffian State for the
5/2 Fractional Quantum Hall Effect: Wavefunction Overlap and
Topological Degeneracy}

\author{Michael. R. Peterson$^1$, Th.~Jolicoeur$^2$, and S. Das
Sarma$^1$}

\affiliation{$^1$Condensed Matter Theory Center, Department of
Physics, University of Maryland, College Park, MD 20742, USA}

\affiliation{$^2$Laboratoire de Physique Th\'eorique et Mod\`eles
Statistiques, Universit\'e Paris-Sud, 91405 Orsay Cedex, France}

\begin{abstract}
We find the finite-width, i.e., the layer thickness, of
experimental quasi-two dimensional systems produces a physical
environment sufficient to stabilize the Moore-Read Pfaffian state
thought to describe the fractional quantum Hall effect at filling
factor $\nu=5/2$.  This conclusion is based on exact calculations
performed in the spherical and torus geometries, studying wavefunction
overlap and ground state degeneracy.

\end{abstract}


\maketitle

{\em Introduction}: Two dimensional (2D) electrons strongly
interacting in the presence of a perpendicular magnetic field
experience the fractional quantum Hall effect~\cite{tsui} (FQHE) at
certain fractional electronic Landau level (LL) filling factors $\nu$
characterized by an incompressible state with fractionally charged
quasiparticles with anyonic, rather than fermionic, statistics, the
observation of which requires clean (high mobility) samples, low
temperatures, and high magnetic fields.  The FQHE abounds in the
lowest Landau level (LLL) with the observation of over 70
odd denominator FQHE states, the most famous being the
Laughlin state~\cite{laugh} describing the FQHE at
$\nu=1/m$ ($m$ odd) -- the odd denominator a consequence of the
Pauli exclusion principle.  We are concerned 
with the FQHE in the second LL (SLL) where the FQHE is 
scarce with only about 8 observed FQHE states which tend to be 
fragile with low activation energies.

The most discussed FQH state in the SLL occurs at the even-denominator
filling factor $\nu=5/2$~\cite{52exp}, thought to be described by the
Moore-Read Pfaffian~\cite{Moore91} state (Pf) which intriguingly possesses
quasiparticle excitations with non-Abelian statistics providing the
tantalizing possibility of topological quantum 
computation~\cite{tqc}.  The presence of this state
challenges our understanding and suggests the condensation of bosons
(perhaps fermion pairs) in a new type of incompressible fluid.
Although the Pf state is the leading candidate for the observed
5/2 FQHE, the {\em actual} nature of the state is currently 
debated~\cite{toke1,toke2,wojs}.  Considering the importance of this 
state, our apparent lack of
understanding of its precise nature, more than 20 years after its
discovery, is both embarrassing and problematic.  This is particularly
true in view of the existence (for more than 15 years) of a beautiful
candidate 5/2 FQHE state, viz. the Pf state~\cite{Moore91}.

The Pf is not as successful in describing the FQHE at
$\nu=5/2$ as the Laughlin theory is in describing the FQHE in the LLL
indicated by the modest overlap between the Pf
wavefunction and the exact Coulomb Hamiltonian
wavefunction~\cite{morf,Rezayi00} (approximately $\sim0.9$ compared to
$\sim 0.999$ for the Laughlin theory wavefunctions).
However, changing~\cite{Rezayi00} the short range components of the Coulomb
interaction can produce an exact wavefunction with near unity overlap
with the Pf.  Furthermore, the actual electron-electron
interaction in the FQHE experimental systems is not purely Coulombic
due to additional physical effects such as disorder, LL mixing,
finite-thickness due to the quasi-2D nature of the system, etc.  A
natural question arises: can any of these effects be
incorporated to produce an exact state that is accurately described by
the Pf wavefunction?  We answer this question
affirmatively with one of the simplest extensions of the pure
Coulombic interaction, namely, the inclusion of finite-thickness
effects.

We find, by including the finite-thickness effects 
perpendicular to the 2D plane, the exact ground state is
very successfully approximated by the Pf model.  We consider two
different complementary compact geometries--the
sphere~\cite{haldane-sphere} and torus~\cite{Haldane85L}.  Throughout
this work we assume the electrons exactly fill half
of the SLL yielding an electron filling factor of $\nu=2+1/2=5/2$ 
(2 coming from completely filling the lowest spin-up and -down bands).
Furthermore, we assume electrons in the SLL to be spin-polarized
since the current consensus supports that conclusion (in any case
the Pf describes a spin-polarized state) and ignore disorder
or LL mixing effects (neglecting LL mixing effects may not be a very
good approximation for the 5/2 FQHE~\cite{dean}).  Hence, the
Hamiltonian is merely the spin-polarized electron interaction
Hamiltonian.

Haldane~\cite{haldane-sphere} showed the Hamiltonian, of
interacting electrons confined in the SLL, can be parameterized by
pseudopotentials $V^{(1)}_m$--the interaction energies
between any pair of electrons with relative angular momentum $m$
\begin{eqnarray}
V^{(1)}_m=\int_{0}^{\infty}dk k [L_1(k^2/2)]^2 L_m(k^2)e^{-k^2}V(k)\;,
\label{vk}
\end{eqnarray}
with $V(k)$ the Fourier transform of the interaction potential and
$L_n(x)$ Laguerre polynomials.  We model the quasi-2D nature of
the experimental system (finite-thickness) by an infinite
square-well potential in the direction perpendicular to the electron
plane, since the best experimental system for the observation of the
5/2 FQHE is typically the GaAs quantum well structure well described
by this model (discussed elsewhere~\cite{longpaper}), given by
\begin{eqnarray}
V(k)=\frac{e^2l}{\epsilon}\frac{1}{k}\frac{\left(3kd+\frac{8\pi^2}{kd}-\frac{32\pi^4(1-\exp(-kd))}{(kd)^2
[(kd)^2+4\pi^2]}\right)}{(kd)^2+4\pi^2}\;,
\end{eqnarray}
where $\epsilon$ is the dielectric constant of the host semiconductor
and $l=\sqrt{\hbar c/eB}$ is the magnetic length.  Eq.~(\ref{vk})
applies to the planar geometry, hence, for the torus it is exact,
however, we also use it on the sphere since (i) it can be argued
it better represents the thermodynamic limit, and (ii) 
convenience.  We do not
expect any qualitative error arising from using these
pseudopotentials for spherical system diagonalization.

{\em Spherical Geometry}: This geometry consists of $N_e$
electrons confined to the spherical surface with a
radial magnetic field produced by a magnetic monopole of strength
$N_\phi/2$ at the center yielding a total magnetic flux piercing the
surface of $N_\phi$ ($N_\phi$ is an integer due to
Dirac's quantization condition).  The total angular momentum $L$ is a
good quantum number and incompressible states are uniform states with
$L=0$ and a non-zero energy gap.  The filling factor 
is $\nu=\lim_{N_e\rightarrow\infty}N_e/N_\phi$.  Using Eq.(~\ref{vk}) 
we calculate entirely within the LLL.  The
relationship between $N_e$ and $N_\phi$ for the
Pf state, describing filling 1/2 in the SLL, is $N_\phi=2N_e-3$
with $-3$ known as the ``shift".

\begin{figure}
\begin{center}
\includegraphics[width=50mm,angle=-90]{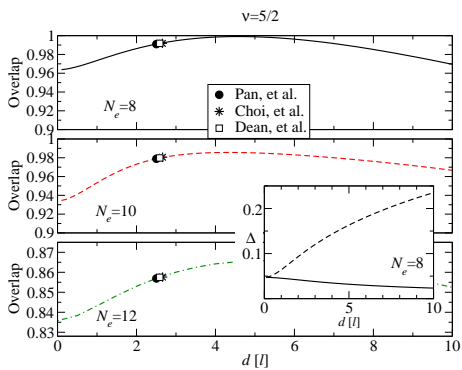}
\caption{(Color online) Wavefunction overlap between the exact ground
state, for a quasi-2D system modeled by an infinite square-well
potential, and the Pf as a function of finite width $d$ for
$N_e=8$, 10, and 12 electrons, respectively.  The
experimental $d$ of Refs.~\onlinecite{pan} (solid
circle),~\onlinecite{choi} (star), and~\onlinecite{dean} (open
square) are also indicated.  The inset shows the 
excitation gap $\Delta$ for $N_e=8$ as a
function of $d$ both in units of $e^2/\epsilon l$ (solid) and
$e^2/\epsilon\sqrt{l^2+d^2}$ (dashed).}
\label{overlaps}
\end{center}
\end{figure}

An appropriate measure to determine the accuracy of the Pf
description of the 5/2 FQHE is the overlap
between the exact ground state and the variational Pf
wavefunctions.  An overlap of unity (zero) indicates the two states
are completely alike (different).  Overlap calculations have been
influential in establishing the nature of the FQHE in the LLL -- in
particular, the primary reason for the theoretical acceptance of the
Laughlin wavefunction as the appropriate description for the 1/3 FQHE
is the large ($>99\%$) overlap it has with exact
finite size numerical many-body wavefunctions.  In the upper, middle,
and lower panels of Fig.~\ref{overlaps} we show the overlap between
the exact ground state for some finite-thickness value $d$ ($=$
quantum well width) and the Pf wavefunction for $N_e=$8, 10, and
12 electrons, respectively (note that the $N_e=$12 system is {\it aliased}
with a FQH state at filling 2/3 and its identification with 1/2 is
dubious~\cite{morf}).  In the zero width case
the overlap is relatively modest but encouraging ($\sim 0.9$).
Surprisingly, the inclusion of finite width causes the overlap to {\em
increase} to a maximum before inevitably 
decreasing to zero for large $d$.  Furthermore, the maximum
occurs at nearly the same value of $d=d_0\sim4 l$ for different system
sizes indicating this conclusion survives in
the thermodynamic limit.  Work by the
authors~\cite{longpaper} showed this effect is true for other
models of finite thickness with a 
similar $d_0$. Thus, the increase of the Pf
overlap with the well width is a generic
qualitative phenomenon, independent of the finite thickness model
employed.  It appears that weakening the Coulomb coupling by
increasing $d$ (to about $4l$) creates an interaction Hamiltonian
favorable to the Pf description.  We mention (emphasized in
Ref.~\onlinecite{longpaper}) that increasing overlap with increasing
layer thickness does {\em not} happen at all for the LLL FQHE where it
is known~\cite{xie} that increasing layer thickness strongly
suppresses the overlap, leading eventually to the destruction of the
FQHE -- e.g., the Laughlin $1/m$ overlap is always maximum at $d=0$.

{\em Torus Geometry}: To test the robustness of this
conclusion we study the Pf state on the
torus.  These results are, in a sense, our main results because
they (i) have less system-size dependence, and (ii) are more general,
i.e., independent of the detailed form of the Pf wavefunction
and dependent only on the topological nature of the underlying 5/2
ground state.  In fact, the finite $d$ spherical geometry
results serve as our motivation and inspiration to
investigate the ground state topological degeneracy on the torus 
at finite $d$, finding the remarkable topological degeneracy -- the
hallmark of a non-Abelian state.

On the torus, there is no ``shift'' in the relation between $N_e$
and $N_\phi$ making a direct comparison possible between various
quantum phases at a given filling factor, such as a Pf state, 
composite fermion (CF)
Fermi sea, or a stripe phase, all possible at $\nu = 1/2$ in the SLL.
These competing states have different spectral signatures
identified by using periodic rectangular geometry with 
sides $a$ and $b$.  The magnetic field prevents 
standard translation operators from commuting, 
however, Haldane~\cite{Haldane85L} showed one can
construct many-body eigenstates with two conserved pseudo-momenta
associated with the translations.  The
two-dimensional pseudo-momentum ${\bf K}$ exist in a Brillouin zone
containing only $N_0^2$ points where $N_0$ is the greatest common
divisor of $N_e$ and $N_\phi$ and $K_x$ ($K_y$) are in
units of $2\pi\hbar/a$ ($2\pi\hbar/b$).  There is an
exact (trivial) degeneracy $q$ due to the center of mass motion 
at filling factor $p/q$ which we ignore as it is unrelated
to the physics (independent of the Hamiltonian).  In the rectangular
unit cell the discrete symmetries relate states at $(\pm K_x,\pm
K_y)$ and as a consequence, we only consider states with
pseudo-momenta in the range $(0\dots N_0/2,0\dots N_0/2)$.

\begin{figure}
\begin{center}
\includegraphics[width=65mm,angle=0]{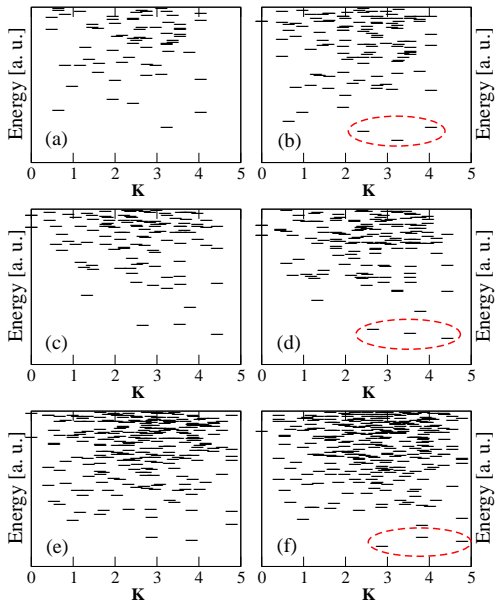}
\caption{Low-lying eigenenergies as a function of the pseudo-momentum
$\sqrt{K_x^2+K_y^2}$ (in physical units) for an aspect ratio of
$a/b=0.75$.  The left panels refer to zero width while the right
panels correspond to the SQ potential of width $d=4 l$ for (a)-(b)
$N_e$=10, (c)-(d) $N_e$=12, (e)-(f) $N_e$=14.}
\label{TorusFig}
\end{center}
\end{figure}

To render the Laughlin state periodic~\cite{Haldane85} the 
recipe is to replace the Jastrow factor $(z_i-z_j)^m$ by a Jacobi
theta function $\theta_1 (z_i-z_j|\tau)^m$ where $\tau =ib/a$ with $z$
the complex electron position.  The Jacobi
theta function has the quasi-periodicity required to construct
a Laughlin state at $\nu =1/m$ observed in numerical
studies~\cite{Haldane85}.  This recipe fails when applied to
the Pf since there are denominators of the form $(z_i-z_j)$
present and the quasi-periodicity of the theta function does not
appear as an overall factor~\cite{Chung07}. The correct
substitution~\cite{Greiter92} is
\begin{equation}
 1/(z_i-z_j)\rightarrow \theta_a
 (z_i-z_j|\tau)/\theta_1(z_i-z_j|\tau), a=2,3,4.\label{PfTorus}
\end{equation}
leading to \textit{three} candidate ground states. This degeneracy is
topological in origin and a signature of the special
properties of the Pf state.  To our knowledge, no
earlier work in the literature has directly discovered this
topological degeneracy on the torus for the 5/2 state in spite of its
great significance.  In the Pf phase of the real system, such as
electrons interacting via Coulomb, we expect the degeneracy
should be approximate for finite size systems and should become
clearer with increasing system size.  Note this trend is
opposite to the overlap trend shown in Fig.~\ref{overlaps} where the
overlap decreases slowly with increasing system size (i.e., from
$N_e=8$ to 12) since the Pf is a variational
approximation.  The wavefunction of the Pf, when written on the
torus using Eq.~\ref{PfTorus}, has pseudo-momenta that are half
reciprocal lattice vectors.  For electrons at $\nu =1/2$ these states
have ${\bf K}=(0,N_0/2),(N_0/2,0),(N_0/2,N_0/2)$.  To explicitly
separate these states we consider the rectangular unit cell (a
hexagonal unit cell has discrete symmetries that render all corners of
the magnetic Brillouin zone equivalent resulting in a trivial Pf
degeneracy).

We have performed exact diagonalizations for $N_e=10,12,14$ electrons
at half filling in the SLL.  Using a pure Coulomb potential ($d=0$) 
we find, for all system sizes, that the spectra are very
sensitive to the unit cell aspect ratio and $N_e$ 
consistent with previous evidence~\cite{Haldane00} for a nearby
compressible stripe phase.  The CF Fermi sea also displays sensitivity
to boundary conditions and changes of the ground state vector ${\bf
K}$ with $N_e$.  This is what we observe for the
same systems with the LLL Coulomb potential at zero width.  No obvious
ground state degeneracy can be discerned in our $d=0$ results.

Switching to nonzero width (using the SQ
potential) we find the appearance of a threefold quasidegenerate set
of states with the right Pf predicted quantum
numbers. This phenomenon is best seen in the region of maximum overlap
found in the spherical geometry, i.e. $d=d_0\sim 4l$, see
Fig.~\ref{TorusFig}.  In this regime, the spectra are much less
sensitive to the aspect ratio, moreover, this behavior is observed
for all finite-width models.  This is the first
time the Pf degeneracy has been observed in an electronic
topological system, i.e., a system described by a two-body
electron-electron interaction Hamiltonian.  For bosons at 
$\nu =1$ with delta function interactions the 
degeneracy appears at $d=0$~\cite{Chung07-2}.

\begin{figure}
\begin{center}
\includegraphics[width=40mm,angle=-90]{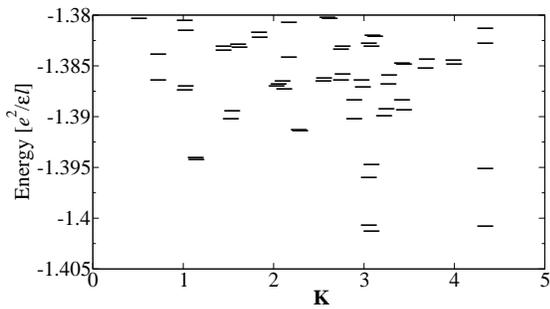}
\caption{Low-lying eigenenergies of $N_e=12$ electrons as a function
{\bf K} using the SQ potential of width $d=4l$ and aspect ratio
$a/b=0.99$. There are doublets at the wavevectors of the Pf
ground states corresponding to the Pfaffian-Anti-Pfaffian symmetric
and antisymmetric combinations.  Due to the almost square symmetry,
there are quasidegeneracies between states at ${\bf K}=(6,0)$ and
${\bf K}=(0,6)$.}
\label{massfig}
\end{center}
\end{figure}

Another feature pointing to the appearance of the Pf is related
to the role of the particle-hole (p-h) symmetry. The Pf
wavefunction is not invariant under this 
symmetry~\cite{Levin07,Lee07} and its p-h conjugate has
been termed the anti-Pfaffian.  While the filling factor $\nu
=1/2$ is p-h invariant, the consequences depend upon the geometry.  On
the sphere, while the Coulomb two-body Hamiltonian has exact p-h
symmetry, the Pf wavefunction has a nontrivial 
shift $-3$ implying that its p-h conjugate requires a different flux 
$N_\phi =2N_e+1$. The zero shift on the torus leads to the
coexistence of these two states, each having exactly the same
threefold topological degeneracy with the same quantum numbers. In a
finite system there is no spontaneous breaking of a discrete symmetry
and we expect tunneling to lead to symmetric and antisymmetric
combinations as eigenstates. We thus expect a doubling of the
Pf states due to the p-h symmetry.  This is best observed
(cf. Fig.~\ref{massfig}) on the torus at finite width with a
nearly square unit cell where the two states at $(0,N_0/2)$ and
$(N_0/2,0)$ (exactly degenerate for the square unit
cell) are very close in energy and all three members
of the Pf multiplet have exactly one partner at a slightly
higher energy. This does not happen at zero width
($d=0$) and is strong evidence for the stabilization of the
Pf physics in the SLL by finite width effects.

The observation of the topological degeneracy on the torus
{\em only} for finite thickness, $d\sim 4l$, precisely where the
overlap is also a maximum on the sphere is, in our opinion, compelling
evidence that the 5/2 FQHE is likely to be a non-Abelian state.

{\em Conclusion}: Our results show the, often assumed 
trivial, effects of the quasi-2D nature of the experimental system
produce an exact state {\em better} described by the
Pf.  The fact that this conclusion is reached in different
finite sized systems for two different geometries (for several
models of thickness) is compelling.  Our results are not
inconsistent with previous work~\cite{toke1,toke2,wojs} in the $d=0$
limit showing the absence of the Pf.  Further, since we find a robust
Pf at finite $d$ the transport gap, seen experimentally, would
be weaker than predicted in $d=0$ theoretical studies since finite
width ``trivially'' reduces energy gaps, see the 
inset of Fig.~\ref{overlaps}.  Thus, the supposed
fragility of the 5/2 state may not necessarily be due to
it being close to a phase boundary, perhaps between a CF Fermi
sea and stripe phase, instead, it may come from the
relatively wide quasi-2D system needed to produce a stable Pf.

In this context, it is useful to mention that although earlier
theoretical work~\cite{morf,Rezayi00} pointed to the importance of
tuning the pseudopotential ratio $V^{(1)}_1/V^{(1)}_3$ in stabilizing
the Pf, finite width affects~\cite{longpaper} {\em all}
pseudopotentials, not just $V^{(1)}_1/V^{(1)}_3$.  Tuning $V^{(1)}_1$
and/or $V^{(1)}_3$, while theoretically 
convenient~\cite{morf,Rezayi00}, is an ambiguous technique
for understanding the stability in real quasi-2D systems where
pseudopotentials cannot be tuned arbitrarily.  Therefore, our work
establishing the optimal stability of the 5/2 Pf at the relatively
large width of $d\sim 4l$,
is important in view of the fact that the Pf is an exact eigenstate
only of a three-body interaction Hamiltonian not expressible in terms
of pseudopotentials.  As shown in Fig.~\ref{overlaps}, the current
quasi-2D samples typically have $d\sim2.5l$ where the wavefunction
overlap is large, yet not optimal, as it would be for thicker samples
with $d\sim4l$.

Our direct numerical finding of the appropriate topological degeneracy
of the 5/2 FQHE state on the torus and the recent experimental
demonstration~\cite{dolev} of the expected $e/4$ quasiparticle charge
in shot noise measurements at $\nu=5/2$, taken together, provide
convincing necessary and sufficient conditions supporting the
contention that the 5/2 FQHE state is indeed the Moore-Read Pfaffian
wavefunction (or some other equivalent state connected adiabatically)
belonging to the (SU(2))$_2$ conformal field theory description, which
obeys the non-Abelian anyonic statistics appropriate for topological
quantum computation~\cite{tqc}, provided the 2D samples are not too
thin.

\begin{acknowledgments}
MRP and SDS acknowledge support from the Microsoft Q Project.
\end{acknowledgments}


\begin{thebibliography}{10}

\bibitem{tsui} D. C. Tsui, H. L. Stormer, and A. C. Gossard,
Phys. Rev. Lett. {\bf 48}, 1559 (1982).

\bibitem{laugh} R. B. Laughlin, Phys. Rev. Lett. {\bf{50}}, 1395
(1983).

\bibitem{52exp} R. Willett {\em{et al}}., Phys. Rev. Lett. {\bf 59}, 1776 (1987).

\bibitem{Moore91} G. Moore and N. Read, Nucl. Phys. B\textbf{360}, 362
(1991).

\bibitem{tqc} S. Das Sarma, M. Freedman, and C. Nayak, Phys.
Rev. Lett. {\bf 94}, 166802 (2005).

\bibitem{toke1} C. Toke and J. K. Jain, Phys. Rev. Lett. {\bf 96},
246805 (2006).

\bibitem{toke2} C. Toke, N. Regnault, and J. K. Jain, Phys. Rev. Lett.
{\bf 98}, 036806 (2007).

\bibitem{wojs} A. Wojs and J. J. Quinn, Phys. Rev. B{\bf 74}, 235319
(2006).

\bibitem{morf} R. H. Morf, Phys. Rev. Lett. {\bf 80}, 1505 (1998).

\bibitem{Rezayi00}
E. H. Rezayi and F. D. M. Haldane, Phys. Rev. Lett. \textbf{84}, 4685
(2000).

\bibitem{haldane-sphere} F. D. M. Haldane, Phys. Rev. Lett. {\bf 51},
605 (1983).

\bibitem{Haldane85L}
F. D. M. Haldane, Phys. Rev. Lett. \textbf{55}, 2095 (1985).

\bibitem{longpaper} M. R. Peterson, T. Jolicoeur, and S. Das Sarma 
(unpublished); see arXiv:0801.4891v1 (2008).

\bibitem{dean} C. R. Dean {\em et al}., Phys. Rev. Lett. \textbf{100}, 
146803 (2008).

\bibitem{xie} S. He {\em et al}., Phys. Rev. B. {\bf 42}, 11376 (1990).

\bibitem{Haldane85} F. D. M. Haldane and E. H. Rezayi, Phys. Rev. B\textbf{31}, 2529
(1985).

\bibitem{Chung07} S. B. Chung and M. Stone, J. Phys. A\textbf{40},
4923 (2007).

\bibitem{Greiter92} M. Greiter, X. G. Wen, and F. Wilczek,
Nucl. Phys. B\textbf{374}, 567 (1992).

\bibitem{Haldane00}
F. D. M. Haldane, E. H. Rezayi, and K. Yang,
Phys. Rev. Lett. \textbf{85}, 5396 (2000).

\bibitem{Chung07-2}
B. Chung and Th. Jolicoeur, arXiv:0712.3185 (2007).

\bibitem{Levin07} M. Levin, B. I. Halperin, and B. Rosenow,
Phys. Rev. Lett. \textbf{99}, 236806 (2007).

\bibitem{Lee07} S.-S. Lee {\em et al}.,
Phys. Rev. Lett. \textbf{99}, 236807 (2007).

\bibitem{dolev} M. Dolev {\em et al}., Nature \textbf{452}, 829 (2008).

\bibitem{pan} W. Pan {\em et al}., Phys. Rev. B \textbf{77}, 075307 (2008).

\bibitem{choi} H. C. Choi {\em et al}., Phys. Rev. B \textbf{77}, 081301(R) (2008).

\end{thebibliography}
\end{document}